\documentclass[prl,reprint,superscriptaddress,
showpacs,amsmath,amssymb,aps]{revtex4-1}
\usepackage{graphicx}
\usepackage{booktabs}
\usepackage{hyperref}

\usepackage{color}

\begin{document}
\title{Direct Detection Constraints on Dark Photons with CDEX-10 Experiment at the China Jinping Underground Laboratory}

\author{Z.~She}
\affiliation{Key Laboratory of Particle and Radiation Imaging (Ministry of Education) and Department of Engineering Physics, Tsinghua University, Beijing 100084}

\author{L.~P.~Jia}
\affiliation{Key Laboratory of Particle and Radiation Imaging (Ministry of Education) and Department of Engineering Physics, Tsinghua University, Beijing 100084}

\author{Q.~Yue}
\altaffiliation [Corresponding author.\newline]
{yueq@mail.tsinghua.edu.cn}
\affiliation{Key Laboratory of Particle and Radiation Imaging (Ministry of Education) and Department of Engineering Physics, Tsinghua University, Beijing 100084}

\author{H.~Ma}
\altaffiliation [Corresponding author.\newline]
{mahao@mail.tsinghua.edu.cn}
\affiliation{Key Laboratory of Particle and Radiation Imaging (Ministry of Education) and Department of Engineering Physics, Tsinghua University, Beijing 100084}

\author{K.~J.~Kang}
\affiliation{Key Laboratory of Particle and Radiation Imaging (Ministry of Education) and Department of Engineering Physics, Tsinghua University, Beijing 100084}

\author{Y.~J.~Li}
\affiliation{Key Laboratory of Particle and Radiation Imaging (Ministry of Education) and Department of Engineering Physics, Tsinghua University, Beijing 100084}

\author{M.~Agartioglu}
\altaffiliation{Participating as a member of TEXONO Collaboration.}
\affiliation{Institute of Physics, Academia Sinica, Taipei 11529}
\affiliation{Department of Physics, Dokuz Eyl\"{u}l University, \.{I}zmir 35160}

\author{H.~P.~An}
\affiliation{Department of Physics, Tsinghua University, Beijing 100084}

\author{J.~P.~Chang}
\affiliation{NUCTECH Company, Beijing 100084}

\author{J.~H.~Chen}
\altaffiliation{Participating as a member of TEXONO Collaboration.}
\affiliation{Institute of Physics, Academia Sinica, Taipei 11529}

\author{Y.~H.~Chen}
\affiliation{YaLong River Hydropower Development Company, Chengdu 610051}

\author{J.~P.~Cheng}
\affiliation{Key Laboratory of Particle and Radiation Imaging (Ministry of Education) and Department of Engineering Physics, Tsinghua University, Beijing 100084}
\affiliation{College of Nuclear Science and Technology, Beijing Normal University, Beijing 100875}

\author{W.~H.~Dai}
\affiliation{Key Laboratory of Particle and Radiation Imaging (Ministry of Education) and Department of Engineering Physics, Tsinghua University, Beijing 100084}

\author{Z.~Deng}
\affiliation{Key Laboratory of Particle and Radiation Imaging (Ministry of Education) and Department of Engineering Physics, Tsinghua University, Beijing 100084}

\author{X.~P.~Geng}
\affiliation{Key Laboratory of Particle and Radiation Imaging (Ministry of Education) and Department of Engineering Physics, Tsinghua University, Beijing 100084}

\author{H.~Gong}
\affiliation{Key Laboratory of Particle and Radiation Imaging (Ministry of Education) and Department of Engineering Physics, Tsinghua University, Beijing 100084}

\author{P.~Gu}
\affiliation{College of Physics, Sichuan University, Chengdu 610064}

\author{Q.~J.~Guo}
\affiliation{School of Physics, Peking University, Beijing 100871}

\author{X.~Y.~Guo}
\affiliation{YaLong River Hydropower Development Company, Chengdu 610051}

\author{L.~He}
\affiliation{NUCTECH Company, Beijing 100084}

\author{S.~M.~He}
\affiliation{YaLong River Hydropower Development Company, Chengdu 610051}

\author{H.~T.~He}
\affiliation{College of Physics, Sichuan University, Chengdu 610064}

\author{J.~W.~Hu}
\affiliation{Key Laboratory of Particle and Radiation Imaging (Ministry of Education) and Department of Engineering Physics, Tsinghua University, Beijing 100084}

\author{T.~C.~Huang}
\affiliation{Sino-French Institute of Nuclear and Technology, Sun Yat-sen University, Zhuhai, 519082}

\author{H.~X.~Huang}
\affiliation{Department of Nuclear Physics, China Institute of Atomic Energy, Beijing 102413}

\author{H.~B.~Li}
\altaffiliation{Participating as a member of TEXONO Collaboration.}
\affiliation{Institute of Physics, Academia Sinica, Taipei 11529}

\author{H.~Li}
\affiliation{NUCTECH Company, Beijing 100084}

\author{J.~M.~Li}
\affiliation{Key Laboratory of Particle and Radiation Imaging (Ministry of Education) and Department of Engineering Physics, Tsinghua University, Beijing 100084}

\author{J.~Li}
\affiliation{Key Laboratory of Particle and Radiation Imaging (Ministry of Education) and Department of Engineering Physics, Tsinghua University, Beijing 100084}

\author{M.~X.~Li}
\affiliation{College of Physics, Sichuan University, Chengdu 610064}

\author{X.~Li}
\affiliation{Department of Nuclear Physics, China Institute of Atomic Energy, Beijing 102413}

\author{X.~Q.~Li}
\affiliation{School of Physics, Nankai University, Tianjin 300071}

\author{Y.~L.~Li}
\affiliation{Key Laboratory of Particle and Radiation Imaging (Ministry of Education) and Department of Engineering Physics, Tsinghua University, Beijing 100084}

\author {B.~Liao}
\affiliation{College of Nuclear Science and Technology, Beijing Normal University, Beijing 100875}

\author{F.~K.~Lin}
\altaffiliation{Participating as a member of TEXONO Collaboration.}
\affiliation{Institute of Physics, Academia Sinica, Taipei 11529}

\author{S.~T.~Lin}
\affiliation{College of Physics, Sichuan University, Chengdu 610064}

\author{S.~K.~Liu}
\affiliation{College of Physics, Sichuan University, Chengdu 610064}

\author {Y.~D.~Liu}
\affiliation{College of Nuclear Science and Technology, Beijing Normal University, Beijing 100875}

\author {Y.~Y.~Liu}
\affiliation{College of Nuclear Science and Technology, Beijing Normal University, Beijing 100875}

\author{Z.~Z.~Liu}
\affiliation{Key Laboratory of Particle and Radiation Imaging (Ministry of Education) and Department of Engineering Physics, Tsinghua University, Beijing 100084}

\author{Y.~C.~Mao}
\affiliation{School of Physics, Peking University, Beijing 100871}

\author{Q.~Y.~Nie}
\affiliation{Key Laboratory of Particle and Radiation Imaging (Ministry of Education) and Department of Engineering Physics, Tsinghua University, Beijing 100084}

\author{J.~H.~Ning}
\affiliation{YaLong River Hydropower Development Company, Chengdu 610051}

\author{H.~Pan}
\affiliation{NUCTECH Company, Beijing 100084}

\author{N.~C.~Qi}
\affiliation{YaLong River Hydropower Development Company, Chengdu 610051}

\author{C.~K.~Qiao}
\affiliation{College of Physics, Sichuan University, Chengdu 610064}

\author{J.~Ren}
\affiliation{Department of Nuclear Physics, China Institute of Atomic Energy, Beijing 102413}

\author{X.~C.~Ruan}
\affiliation{Department of Nuclear Physics, China Institute of Atomic Energy, Beijing 102413}

\author{B.~Sevda}
\altaffiliation{Participating as a member of TEXONO Collaboration.}
\affiliation{Institute of Physics, Academia Sinica, Taipei 11529}
\affiliation{Department of Physics, Dokuz Eyl\"{u}l University, \.{I}zmir 35160}

\author{C.~S.~Shang}
\affiliation{YaLong River Hydropower Development Company, Chengdu 610051}

\author{V.~Sharma}
\altaffiliation{Participating as a member of TEXONO Collaboration.}
\affiliation{Institute of Physics, Academia Sinica, Taipei 11529}
\affiliation{Department of Physics, Banaras Hindu University, Varanasi 221005}

\author{L.~Singh}
\altaffiliation{Participating as a member of TEXONO Collaboration.}
\affiliation{Institute of Physics, Academia Sinica, Taipei 11529}
\affiliation{Department of Physics, Banaras Hindu University, Varanasi 221005}

\author{M.~K.~Singh}
\altaffiliation{Participating as a member of TEXONO Collaboration.}
\affiliation{Institute of Physics, Academia Sinica, Taipei 11529}
\affiliation{Department of Physics, Banaras Hindu University, Varanasi 221005}

\author {T.~X.~Sun}
\affiliation{College of Nuclear Science and Technology, Beijing Normal University, Beijing 100875}

\author{C.~J.~Tang}
\affiliation{College of Physics, Sichuan University, Chengdu 610064}

\author{W.~Y.~Tang}
\affiliation{Key Laboratory of Particle and Radiation Imaging (Ministry of Education) and Department of Engineering Physics, Tsinghua University, Beijing 100084}

\author{Y.~Tian}
\affiliation{Key Laboratory of Particle and Radiation Imaging (Ministry of Education) and Department of Engineering Physics, Tsinghua University, Beijing 100084}

\author {G.~F.~Wang}
\affiliation{College of Nuclear Science and Technology, Beijing Normal University, Beijing 100875}

\author{L.~Wang}
\affiliation{Department of Physics, Beijing Normal University, Beijing 100875}

\author{Q.~Wang}
\affiliation{Key Laboratory of Particle and Radiation Imaging (Ministry of Education) and Department of Engineering Physics, Tsinghua University, Beijing 100084}
\affiliation{Department of Physics, Tsinghua University, Beijing 100084}

\author{Y.~Wang}
\affiliation{Key Laboratory of Particle and Radiation Imaging (Ministry of Education) and Department of Engineering Physics, Tsinghua University, Beijing 100084}
\affiliation{Department of Physics, Tsinghua University, Beijing 100084}

\author{Y.~X.~Wang}
\affiliation{School of Physics, Peking University, Beijing 100871}

\author{Z.~Wang}
\affiliation{College of Physics, Sichuan University, Chengdu 610064}

\author{H.~T.~Wong}
\altaffiliation{Participating as a member of TEXONO Collaboration.}
\affiliation{Institute of Physics, Academia Sinica, Taipei 11529}

\author{S.~Y.~Wu}
\affiliation{YaLong River Hydropower Development Company, Chengdu 610051}

\author{H.~Y.~Xing}
\affiliation{College of Physics, Sichuan University, Chengdu 610064}

\author{Y.~Xu}
\affiliation{School of Physics, Nankai University, Tianjin 300071}

\author{T.~Xue}
\affiliation{Key Laboratory of Particle and Radiation Imaging (Ministry of Education) and Department of Engineering Physics, Tsinghua University, Beijing 100084}

\author{Y.~L.~Yan}
\affiliation{College of Physics, Sichuan University, Chengdu 610064}

\author{L.~T.~Yang}
\altaffiliation [Corresponding author.\newline]
{yanglt@mail.tsinghua.edu.cn}
\affiliation{Key Laboratory of Particle and Radiation Imaging (Ministry of Education) and Department of Engineering Physics, Tsinghua University, Beijing 100084}

\author{N.~Yi}
\affiliation{NUCTECH Company, Beijing 100084}

\author{C.~X.~Yu}
\affiliation{School of Physics, Nankai University, Tianjin 300071}

\author{H.~J.~Yu}
\affiliation{NUCTECH Company, Beijing 100084}

\author{J.~F.~Yue}
\affiliation{YaLong River Hydropower Development Company, Chengdu 610051}

\author{M.~Zeng}
\affiliation{Key Laboratory of Particle and Radiation Imaging (Ministry of Education) and Department of Engineering Physics, Tsinghua University, Beijing 100084}

\author{Z.~Zeng}
\affiliation{Key Laboratory of Particle and Radiation Imaging (Ministry of Education) and Department of Engineering Physics, Tsinghua University, Beijing 100084}

\author{B.~T.~Zhang}
\affiliation{Key Laboratory of Particle and Radiation Imaging (Ministry of Education) and Department of Engineering Physics, Tsinghua University, Beijing 100084}

\author{L.~Zhang}
\affiliation{College of Physics, Sichuan University, Chengdu 610064}

\author {F.~S.~Zhang}
\affiliation{College of Nuclear Science and Technology, Beijing Normal University, Beijing 100875}

\author{Z.~Y.~Zhang}
\affiliation{Key Laboratory of Particle and Radiation Imaging (Ministry of Education) and Department of Engineering Physics, Tsinghua University, Beijing 100084}

\author{M.~G.~Zhao}
\affiliation{School of Physics, Nankai University, Tianjin 300071}

\author{J.~F.~Zhou}
\affiliation{YaLong River Hydropower Development Company, Chengdu 610051}

\author{Z.~Y.~Zhou}
\affiliation{Department of Nuclear Physics, China Institute of Atomic Energy, Beijing 102413}

\author{J.~J.~Zhu}
\affiliation{College of Physics, Sichuan University, Chengdu 610064}

\collaboration{CDEX Collaboration}
\noaffiliation
\date{\today}
\begin{abstract}
	We report constraints on the dark photon effective kinetic mixing parameter (${\kappa}$) with data taken from two ${p}$-type point-contact germanium detectors of the CDEX-10 experiment at the China Jinping Underground Laboratory. 
	The 90\% confidence level upper limits on ${\kappa}$ of solar dark photon from 205.4 kg-day exposure are derived, probing new parameter space with masses (${m_V}$) from 10 to 300 eV/${c^2}$ in direct detection experiments. 
	Considering dark photon as the cosmological dark matter, limits at 90\% confidence level with ${m_V}$ from 0.1 to 4.0 keV/${c^2}$ are set from 449.6 kg-day data, with a minimum of ${\rm{\kappa=1.3 \times 10^{-15}}}$ at ${\rm{m_V=200\ eV/c^2}}$. 
\end{abstract}

\maketitle 
\paragraph{}
\textit{Introduction.--}Extensive cosmological and astrophysical observations at different scales provide compelling evidence on the existence of missing matter in the Universe \cite{Young2017}. 
This motivates intense efforts to study extensions to the standard model (SM) \cite{Olive2014}.
An attractive class introduces the dark photon (denoted as ${\gamma^*}$ in this Letter) with mass (${m_V}$) of keV/${c^2}$ scale as a potential hidden particle as a dark matter (DM) candidate \cite{Boehm2004, Pospelov2008}.
The ${\gamma^*}$ can, in addition, be a new interaction mediator between DM and the SM particles \cite{Jaeckel2010,Jaeckel2013A}. 
The properties of ${\gamma^*}$, and in particular its interaction with SM particles, are parametrized by its effective kinetic mixing parameter (${\kappa}$) with the SM photons \cite{Holdom1986}.

\paragraph{}
Intense gamma sources therefore provide an excellent platform to look for ${\gamma^*}$. 
Searches are conducted using high-power laser \cite{Ahlers2008,Schwarz2015} or helioscopes \cite{Redondo2008} or via the studies of photons from stars \cite{An2013New} and cosmic microwave background \cite{Jaeckel2010, Mirizzi2009}.
Other studies are based on data taken with underground DM and neutrino experiments \cite{Aguilar2019,Angloher2017,Gninenko2008, An2015, An2013Dark, Bloch2017, XENON1t, Abramoff2019}. 

\paragraph{} 
\textit{CDEX-10 experiment.--}The second phase of the China Dark Matter Experiment (CDEX-10), with scientific goals of searching for light DM, is based on the 10-kg ${p}$-type point contact germanium (PPCGe) detector array at the China Jinping Underground Laboratory (CJPL) with a rock overburden of 2400 meters  \cite{Kang2013,Cheng2017,Wu2013,Yang2019,Liu2019}. 
The detector array, consisting of three triple-element PPCGe detector strings (named C10-A, B, C, respectively), surrounded by the 20 cm thick high-purity oxygen-free copper as a passive shield against ambient radioactivity, is directly immersed in liquid nitrogen (${\rm{LN_2}}$) for cooling. 
The ${\rm{LN_2}}$ cryostat, with an outer diameter of 1.5 m and a height of 1.9 m, together with the electronics readout and the data acquisition (DAQ) systems, is placed inside a shielding with 1 m thick polyethylene walls at CJPL-I \cite{Jiang2018,Jiang2019}. 
The DAQ system consists of five readout channels with specific shaping and gain parameters for each PPCGe detector to record signal information. 
The configuration of the detector system was described in detail previously \cite{Jiang2018, Jiang2019}. 
In this Letter, a raw dataset with exposure time of 363.9 day from two of the CDEX-10 detectors, named, C10-B1 and C10-C1, are used for ${\gamma^*}$ analysis.

\paragraph{}
Data taking with C10-B1 and  C10-C1 was performed from February 2017 to August 2018. Independent data-quality filters were carried out for the two detectors. 
The resulting data with good qualities have 234.6 running days for C10-B1 and 276.8 running days for C10-C1, respectively. 
The dead times of the DAQ plus the veto time due to the reset signals of the preamplifiers were measured by the random trigger events to be 5.7\% and 5.5\% of the survival data of the two detectors, respectively.
Energy calibrations were done by the zero energy (defined by the random trigger events) and the internal cosmogenic ${K}$-shell x-ray peaks: 8.98 keVee of ${^{65}}$Zn and 10.37 keVee of ${^{68,71}}$Ge.  
The data preprocessing procedures for a single detector were described in previous papers \cite{Jiang2018,Jiang2019}, including energy calibrations, pedestal cut, physics event selection, and bulk or surface event discrimination \cite{Yang2018Bulk}.
The live times were reduced by 1.1\% for C10-B1 and 0.6\% for C10-C1 after pedestal cuts, which were energy independent and measured by the survival of the random trigger events. 
The residual cross talks between the two detectors were studied by the trigger time information. 
Coincident events in both detectors are extremely few in the energy region of interests and can be negligible for this study.
The dead layer thicknesses of these two detectors are set to be ${\rm{0.88\pm0.12\ mm}}$, considering the same crystal size, crystal structure and fabrication procedure as the CDEX-1B detector \cite{Yang2018}.
This gives rise to a fiducial mass of 939 g and, accordingly, valid physics data of 205.4 and 244.2 kg-day for C10-B1 and C10-C1, respectively. 

\begin{figure}
	\includegraphics[width = .5\textwidth]{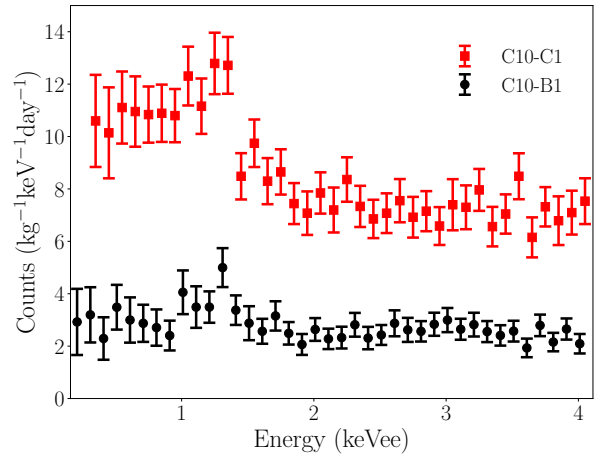}
	\caption{The measured energy spectra of C10-B1 and C10-C1 germanium detectors at exposure of 205.4 kg-day and 244.2 kg-day, respectively.}
	\label{EnergySpectrum}
\end{figure}

\paragraph{}
The energy spectra of these detectors for ${\gamma^*}$ analysis are shown in Fig. \ref{EnergySpectrum}. 
The analysis thresholds are 160 eVee (``eVee" represents electron equivalent energy)  for C10-B1 and 300 eVee for C10-C1.
The energy peaks around 1.2 keVee are originated from the ${L}$-shell x rays of the internal cosmogenic radionuclides. 
The event rates are about 2.5 and 7.6 ${\rm{counts\ kg^{-1}\ keVee^{-1}\ day^{-1}}}$ in the 2${-}$4 keVee energy range for these two detectors. 
The higher analysis threshold and event rate make the C10-C1 less sensitive to the study of solar ${\gamma^*}$, such that only 205.4 kg-day data of C10-B1 are used in this analysis. 
For ${\gamma^*}$ DM analysis, however, data of these two detectors are combined to produce a larger exposure of 449.6 kg-day.

\paragraph{} 
\textit{Data analysis.--}A minimal-${\chi^2}$ analysis method, following our previous WIMPs analysis \cite{Jiang2018, Yue2014}, is applied to probe (${m_V}$,${\kappa}$), in which ${\chi^2}$ is defined as
\begin{equation}
\label{chi2}
\chi^2(m_V,\kappa) = \sum_{j=1,2}\sum^{N_j}_{i=0}{\frac{[n_{i,j}-S_i(m_V,\kappa)-B_{i,j}]^2}{\sigma_{stat,i, j}^2+\sigma_{syst,i,j}^2}},
\end{equation}
where ${n_{i,j}}$ is the measured event rate corresponding to the ${i}$th energy bin and the ${j}$th detector (${j=1}$ for C10-B1 and ${j=2}$ for C10-C1), and ${S_i(m_V,\kappa)}$ is the expected rate at certain ${m_V}$ and ${\kappa}$. The uncertainties ${\sigma}$ include both statistical (stat) and systematical (syst) components denoted as subscripts.
${B_{i,j}}$ denotes the background contribution assuming for C10-B1 to be a flat continuum plus several known background ${L}$-shell x-ray peaks, including  ${^{68}}$Ge (1.298 keV), ${^{68}}$Ga (1.194 keV), ${^{65}}$Zn (1.096 keV), ${^{54}}$Mn (0.695 keV), and ${^{49}}$V (0.564 keV), for C10-B1 \cite{Jiang2018}. 
The intensities of these peaks from ${L}$-shell x rays are derived from those of measured ${K}$-shell x-ray peaks at a higher energy range of 4${-}$12 keVee corrected by the known K/L ratios \cite{Bahcall1963, Jiang2018}. 
However, the background form for C10-C1 is slightly different, so that the flat continuum profile is replaced by a linear one.

\paragraph{}
The best estimator of ${\kappa}$ (denoted with ${\hat{\kappa}}$) at certain ${m_V}$ is evaluated by minimizing the ${\chi^2}$ values, from which upper limits at 90\% confidence level (C.L.) are derived \cite{Feldman1998}.
The background contributions ${B_{i,j}}$ are restricted to be positive definite.

\begin{figure}
	\includegraphics[width = .5\textwidth]{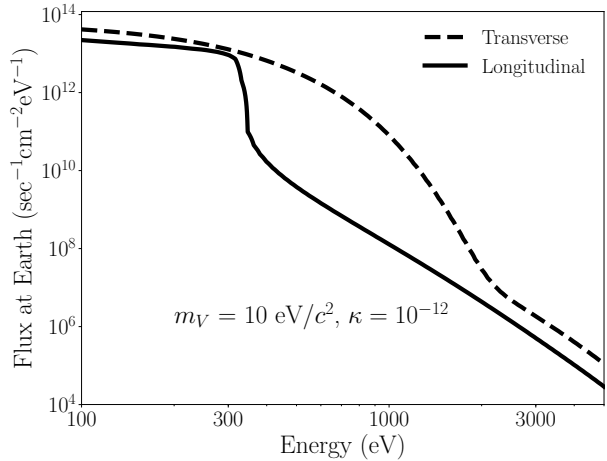}
	\caption{The differential fluxes of solar ${\gamma^*}$ on Earth originated in the Stueckelberg case. The dashed line represents the transverse ${\gamma^*}$, while the longitudinal one is shown as a solid line.}
	\label{diffFlux}
\end{figure}

\paragraph{} 
\textit{Solar dark photon.--}The most significant gamma source seen on Earth is the Sun and solar ${\gamma^*}$ can be probed in underground low-background experiments and cosmological observations. 
The solar ${\gamma^*}$ flux received on Earth can be determined following the procedures proposed in Refs. \cite{An2013New, An2013Dark, Horvat2013}. 
In this study it is assumed that the solar ${\gamma^*}$ originates from the Stueckelberg case with nondynamical mass, and its differential flux spectra in polarization functions are shown in Fig. \ref{diffFlux} at certain (${m_V,\ \kappa}$). 
Experimental signatures of ${\gamma^*}$ are through their absorption and conversion to electrons in the detectors with a process analogous to the photoelectric effect. 
Consequently, the absorption rates of ${\gamma^*}$ inside a PPCGe detector are related to the photo-absorption cross sections in the germanium crystal. 
The energy is deposited locally due to the short range of ionized electrons inside PPCGe detectors. 
The expected energy spectrum of the solar ${\gamma^*}$ in a germanium detector is described by \cite{An2013Dark} 
\begin{equation}
\label{CountSolarDP}
\frac{dR(E)}{dE}= \frac{M}{\rho}\frac{E}{m_V}\left(\frac{d\Phi_T}{dE}\Gamma_T+\frac{d\Phi_L}{dE}\Gamma_L\right)B_R,
\end{equation}
where ${M}$ and ${\rho}$ represent the mass and density of the germanium crystal, respectively, ${E}$ is the energy of ${\gamma^*}$, ${\Gamma_{T,L}(m_V,\kappa)}$ denote for the absorption rates for the ${\gamma^*}$ in transverse and longitudinal polarizations, ${\frac{d\Phi_{T,L}}{dE}(m_V, \kappa)}$ are the differential fluxes on Earth with the different polarization functions mentioned above, while ${B_R}$ represents the branching ratio of the photoionization to the total absorption and it is equal to 1 for ${m_V < 300\ \rm{eV}/}$${c^2}$ \cite{An2013Dark}. 
We neglect for simplicity the anisotropy of the germanium crystal in this analysis. 
The polarization functions for isotropic nonmagnetic materials can be derived from their optical refraction indices \cite{Henke1993}, based on which the absorption rates of transverse and longitudinal ${\gamma^*}$ are calculated according to the procedures described in Refs. \cite{An2013Dark,Horvat2013,An2013New}. 

\begin{figure}
	\centering
	\includegraphics[width = .5\textwidth]{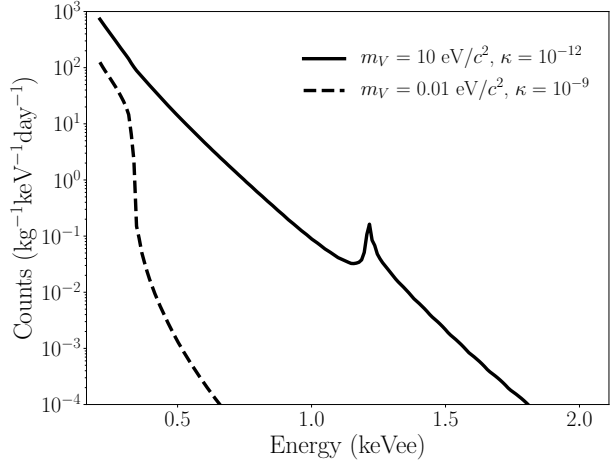}
	\caption{The expected energy spectra of the solar ${\gamma^*}$ at the specified (${m_V,\ \kappa}$) in the CDEX-10 experiment.}
	\label{signature}
\end{figure}

\paragraph{}
Substituting the absorption rates and the differential fluxes of the solar ${\gamma^*}$ into Eq. \eqref{CountSolarDP}, one gets the expected energy spectrum of the solar ${\gamma^*}$ in a germanium detector. 
Folding in the energy resolution of the detector, the expected energy spectra at two different (${m_V,\ \kappa}$) values are depicted in Fig. \ref{signature} as illustration. 

\begin{figure}
	\centering
	\includegraphics[width = .5\textwidth]{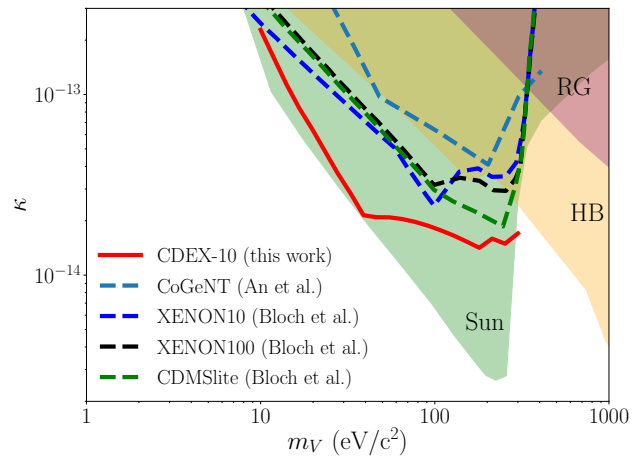}
	\caption{Upper limits (90\% C.L.) of ${\kappa}$ for solar ${\gamma^*}$ as a function of ${m_V}$. Results derived from direct detection experiments, i.e. CoGeNT \cite{An2013Dark}, XENON and CDMSLite \cite{Bloch2017}, are shown as dashed lines. Shaded regions are the results from the astrophysical constraints including the Sun, horizontal branch stars (HB), and the red giant (RG) \cite{Gondolo2009,An2013Dark,Redondo2013}.}
	\label{SDPExclusive}	
\end{figure}

\paragraph{} 
No signals on solar ${\gamma^*}$ are observed with ${\kappa}$ at 1 ${\sigma}$ excess from null effects for ${m_V \sim}$ ${10-300}$ eV/${c^2}$.
Upper limits on ${\kappa}$ at 90\% C.L. are displayed in Fig. \ref{SDPExclusive}, superimposed with previous results from direct detection experiments and astrophysical observations.
The constraints from this work are the most stringent in ${m_V}$ from 10 to 300 eV/${c^2}$, among the direct detection experiments to date. 

\paragraph{}
The enhanced sensitivities of our measurement can be understood as follows.
Experimental sensitivities are determined by the absorption rates ${\Gamma_{T,L}(m_V,\kappa)}$ of Eq. \eqref{CountSolarDP}. 
At ${E\sim}$ ${O}$(10 eV) and ${m_V \sim}$ ${O}$(10 eV/${c^2}$), the absorption rates of germanium detectors is larger than xenon detectors for longitudinal solar ${\gamma^*}$, while the differences between the two nuclei are minor for transverse solar ${\gamma^*}$. 
In addition, the longitudinal ${\gamma^*}$ dominates the solar flux at ${m_V \le}$ ${O}$(10 eV/${c^2}$), while the transverse component is larger at ${m_V \ge}$ ${O}$(10 eV/${c^2}$) \cite{Bloch2017, An2013Dark}. 
When ${m_V}$ increases, the main component of solar ${\gamma^*}$ evolves from longitudinal into transverse polarization. 
Consequently, C10-B1 with 160 eVee energy threshold has better sensitivity on both longitudinal and transverse ${\gamma^*}$, while the CoGeNT data \cite{Aalseth2011} at 450 eV threshold are not sensitive to the longitudinal component. 

\begin{figure}
	\centering
	\includegraphics[width = .5\textwidth]{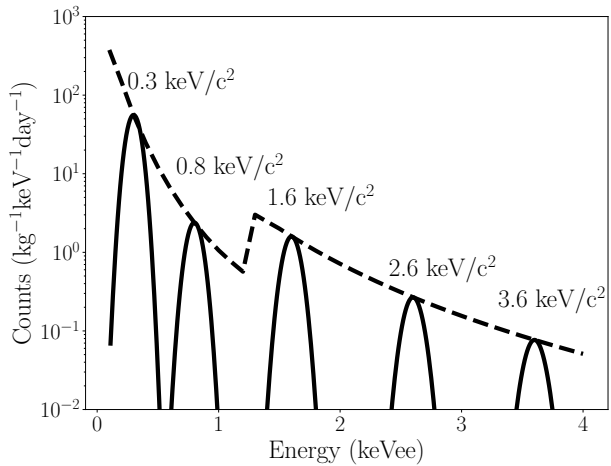}
	\caption{The expected measurable energy spectra of ${\gamma^*}$ DM at the C10-B1 detector with specific ${m_V}$ at ${\rm{\kappa=10^{-14}}}$. The profile of energy peaks is depicted by a dashed line, similar to a photon-electric cross section curve.}
	\label{DPDM}
\end{figure}

\paragraph{}
\textit{Dark photon dark matter.--}The ${\gamma^*}$ DM is also a possible contributor in experiments on rare event searches. 
Considering their number density and local velocity under DM halo assumption \cite{Navarro1996}, the kinetic energy of ${\gamma^*}$ DM is negligible compared with ${m_V}$ in the calculation of energy deposition.
Using the same absorption rate of ${\gamma^*}$ DM derived in Ref. \cite{An2013New}, the expected differential event rates in a detector can be calculated with Eq. \eqref{DPDMFlux} \cite{An2015}, and the expected spectra at specific ${m_V}$ are shown in Fig. \ref{DPDM} as illustration. 
The differential rate observable at detectors is
\begin{equation}
\begin{aligned}
\frac{dR}{dE}=
V\rho_\chi \kappa^2 I_{r},
\end{aligned}
\label{DPDMFlux}
\end{equation}

where ${V}$ is the volume of a detector, ${\rho_\chi}$ denotes for the local DM density (taken to be ${\rm{0.3\ GeV/cm^3}}$ in this Letter) and ${I_{r}}$ represents the imaginary part of the square of the refraction index for certain material tabulated in Ref. \cite{Henke1993}. 

\begin{figure}
	\centering
	\includegraphics[width = .5\textwidth]{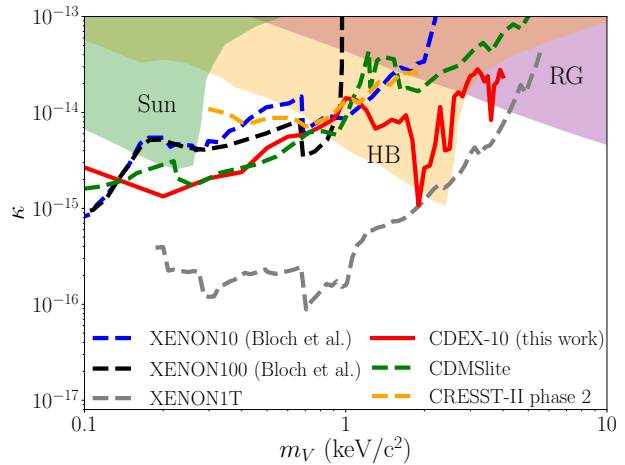}
	\caption{Upper limits with 90\% C.L. on ${\kappa}$ of ${\gamma^*}$ DM (solid red line) superimposed with the results from other direct DM searches (dash line). The constraints from CRESST-II phase-2 \cite{Angloher2017}, XENON-10, XENON-100, CDMSlite \cite{Bloch2017}, and XENON1T \cite{XENON1t} are shown in orange, blue, black, green, and gray lines, respectively. The astrophysical constraints from anomalous energy loss in the Sun, horizontal branch stars, and red giant stars are also shown as shaded regions \cite{An2015}.}
	\label{DPDMExclusive}
\end{figure}

\paragraph{}
Following similar ${\chi^2}$ minimization analysis discussed above, no significant signal of ${\gamma^*}$ DM is observed. 
The upper limits on ${\kappa}$ at 90\% C.L. from ${0.1}$ to ${4.0}$ keV/${c^2}$ are determined and depicted in Fig. \ref{DPDMExclusive}, superimposed with those from previous direct detection experiments and astrophysical observations. 
In particular,  the most stringent constraint is at ${m_V}$ = 200 eV/${c^2}$ where ${\kappa = 1.3 \times10^{-15}}$. 
The complexities of the exclusion curve above 1 keVee can be attributed to the ${L}$-shell x rays of cosmogenic radionuclides such as ${^{68}}$Ge and ${^{68}}$Ga.

\paragraph{} 
\textit{Conclusion.--}In this Letter, we report the results on the searches of ${\gamma^*}$ based on the data from the CDEX-10 experiment, with exposures of 205.4 kg-day from the C10-B1 detector and 244.2 kg-day from the C10-C1 detector. 
A multidetector joint analysis is implemented for a ${\gamma^*}$ DM search.
No significant signal is found and the most stringent direct detection limits on ${\kappa}$ are set for solar ${\gamma^*}$ with ${m_V}$ of 10 to 300 eV/${c^2}$. 
New parameter space of ${\kappa}$ for ${\gamma^*}$ DM is probed within the 0.1${-}$4.0 keV/${c^2}$ mass region as well.

\paragraph{}
This work was supported by the National Key Research and Development Program of China (Grant No. 2017YFA0402200) and the National Natural Science Foundation of China (Grants No. 11725522, No. 11675088, and No. 11475099) and the Tsinghua University Initiative Scientific Research Program (Grant No. 20197050007). 
\paragraph{}
Z. She and L.P. Jia contributed equally to this work.

\bibliographystyle{apsrev4-1} 
\bibliography{darkphoton.bib}

\begin{thebibliography}{38}%
\makeatletter
\providecommand \@ifxundefined [1]{%
 \@ifx{#1\undefined}
}%
\providecommand \@ifnum [1]{%
 \ifnum #1\expandafter \@firstoftwo
 \else \expandafter \@secondoftwo
 \fi
}%
\providecommand \@ifx [1]{%
 \ifx #1\expandafter \@firstoftwo
 \else \expandafter \@secondoftwo
 \fi
}%
\providecommand \natexlab [1]{#1}%
\providecommand \enquote  [1]{``#1''}%
\providecommand \bibnamefont  [1]{#1}%
\providecommand \bibfnamefont [1]{#1}%
\providecommand \citenamefont [1]{#1}%
\providecommand \href@noop [0]{\@secondoftwo}%
\providecommand \href [0]{\begingroup \@sanitize@url \@href}%
\providecommand \@href[1]{\@@startlink{#1}\@@href}%
\providecommand \@@href[1]{\endgroup#1\@@endlink}%
\providecommand \@sanitize@url [0]{\catcode `\\12\catcode `\$12\catcode
  `\&12\catcode `\#12\catcode `\^12\catcode `\_12\catcode `\%12\relax}%
\providecommand \@@startlink[1]{}%
\providecommand \@@endlink[0]{}%
\providecommand \url  [0]{\begingroup\@sanitize@url \@url }%
\providecommand \@url [1]{\endgroup\@href {#1}{\urlprefix }}%
\providecommand \urlprefix  [0]{URL }%
\providecommand \Eprint [0]{\href }%
\providecommand \doibase [0]{http://dx.doi.org/}%
\providecommand \selectlanguage [0]{\@gobble}%
\providecommand \bibinfo  [0]{\@secondoftwo}%
\providecommand \bibfield  [0]{\@secondoftwo}%
\providecommand \translation [1]{[#1]}%
\providecommand \BibitemOpen [0]{}%
\providecommand \bibitemStop [0]{}%
\providecommand \bibitemNoStop [0]{.\EOS\space}%
\providecommand \EOS [0]{\spacefactor3000\relax}%
\providecommand \BibitemShut  [1]{\csname bibitem#1\endcsname}%
\let\auto@bib@innerbib\@empty
\bibitem [{\citenamefont {Young}(2017)}]{Young2017}%
  \BibitemOpen
  \bibfield  {author} {\bibinfo {author} {\bibfnamefont {B.~L.}\ \bibnamefont
  {Young}},\ }\href {\doibase 10.1007/s11467-016-0583-4} {\bibfield  {journal}
  {\bibinfo  {journal} {Front. Phys.}\ }\textbf {\bibinfo {volume} {12}},\
  \bibinfo {pages} {121201} (\bibinfo {year} {2017})}\BibitemShut {NoStop}%
\bibitem [{\citenamefont {Olive}\ \emph {et~al.}(2014)\citenamefont {Olive}
  \emph {et~al.}}]{Olive2014}%
  \BibitemOpen
  \bibfield  {author} {\bibinfo {author} {\bibfnamefont {K.~A.}\ \bibnamefont
  {Olive}} \emph {et~al.} (\bibinfo {collaboration} {Particle Data Group}),\
  }\href {\doibase 10.1088/1674-1137/38/9/090001} {\bibfield  {journal}
  {\bibinfo  {journal} {Chin. Phys. C}\ }\textbf {\bibinfo {volume} {38}},\
  \bibinfo {pages} {090001} (\bibinfo {year} {2014})}\BibitemShut {NoStop}%
\bibitem [{\citenamefont {Boehm}\ and\ \citenamefont
  {Fayet}(2004)}]{Boehm2004}%
  \BibitemOpen
  \bibfield  {author} {\bibinfo {author} {\bibfnamefont {C.}~\bibnamefont
  {Boehm}}\ and\ \bibinfo {author} {\bibfnamefont {P.}~\bibnamefont {Fayet}},\
  }\href {\doibase 10.1016/j.nuclphysb.2004.01.015} {\bibfield  {journal}
  {\bibinfo  {journal} {Nucl. Phys. B}\ }\textbf {\bibinfo {volume} {683}},\
  \bibinfo {pages} {219 } (\bibinfo {year} {2004})}\BibitemShut {NoStop}%
\bibitem [{\citenamefont {Pospelov}\ \emph {et~al.}(2008)\citenamefont
  {Pospelov}, \citenamefont {Ritz},\ and\ \citenamefont
  {Voloshin}}]{Pospelov2008}%
  \BibitemOpen
  \bibfield  {author} {\bibinfo {author} {\bibfnamefont {M.}~\bibnamefont
  {Pospelov}}, \bibinfo {author} {\bibfnamefont {A.}~\bibnamefont {Ritz}}, \
  and\ \bibinfo {author} {\bibfnamefont {M.}~\bibnamefont {Voloshin}},\ }\href
  {\doibase 10.1103/PhysRevD.78.115012} {\bibfield  {journal} {\bibinfo
  {journal} {Phys. Rev. D}\ }\textbf {\bibinfo {volume} {78}},\ \bibinfo
  {pages} {115012} (\bibinfo {year} {2008})}\BibitemShut {NoStop}%
\bibitem [{\citenamefont {Jaeckel}\ and\ \citenamefont
  {Ringwald}(2010)}]{Jaeckel2010}%
  \BibitemOpen
  \bibfield  {author} {\bibinfo {author} {\bibfnamefont {J.}~\bibnamefont
  {Jaeckel}}\ and\ \bibinfo {author} {\bibfnamefont {A.}~\bibnamefont
  {Ringwald}},\ }\href {\doibase 10.1146/annurev.nucl.012809.104433} {\bibfield
   {journal} {\bibinfo  {journal} {Annu. Rev. Nucl. Part. Sci.}\ }\textbf
  {\bibinfo {volume} {60}},\ \bibinfo {pages} {405} (\bibinfo {year}
  {2010})}\BibitemShut {NoStop}%
\bibitem [{\citenamefont {Jaeckel}(2012)}]{Jaeckel2013A}%
  \BibitemOpen
  \bibfield  {author} {\bibinfo {author} {\bibfnamefont {J.}~\bibnamefont
  {Jaeckel}},\ }\href@noop {} {\bibfield  {journal} {\bibinfo  {journal}
  {Frascati Phys. Ser.}\ }\textbf {\bibinfo {volume} {56}},\ \bibinfo {pages}
  {172} (\bibinfo {year} {2012})}\BibitemShut {NoStop}%
\bibitem [{\citenamefont {Holdom}(1986)}]{Holdom1986}%
  \BibitemOpen
  \bibfield  {author} {\bibinfo {author} {\bibfnamefont {B.}~\bibnamefont
  {Holdom}},\ }\href {\doibase 10.1016/0370-2693(86)91377-8} {\bibfield
  {journal} {\bibinfo  {journal} {Phys. Lett. B}\ }\textbf {\bibinfo {volume}
  {166}},\ \bibinfo {pages} {196 } (\bibinfo {year} {1986})}\BibitemShut
  {NoStop}%
\bibitem [{\citenamefont {Ahlers}\ \emph {et~al.}(2008)\citenamefont {Ahlers},
  \citenamefont {Gies}, \citenamefont {Jaeckel}, \citenamefont {Redondo},\ and\
  \citenamefont {Ringwald}}]{Ahlers2008}%
  \BibitemOpen
  \bibfield  {author} {\bibinfo {author} {\bibfnamefont {M.}~\bibnamefont
  {Ahlers}}, \bibinfo {author} {\bibfnamefont {H.}~\bibnamefont {Gies}},
  \bibinfo {author} {\bibfnamefont {J.}~\bibnamefont {Jaeckel}}, \bibinfo
  {author} {\bibfnamefont {J.}~\bibnamefont {Redondo}}, \ and\ \bibinfo
  {author} {\bibfnamefont {A.}~\bibnamefont {Ringwald}},\ }\href {\doibase
  10.1103/PhysRevD.77.095001} {\bibfield  {journal} {\bibinfo  {journal} {Phys.
  Rev. D}\ }\textbf {\bibinfo {volume} {77}},\ \bibinfo {pages} {095001}
  (\bibinfo {year} {2008})}\BibitemShut {NoStop}%
\bibitem [{\citenamefont {Schwarz}\ \emph {et~al.}(2015)\citenamefont
  {Schwarz}, \citenamefont {Knabbe}, \citenamefont {Lindner}, \citenamefont
  {Redondo}, \citenamefont {Ringwald}, \citenamefont {Schneide}, \citenamefont
  {Susol},\ and\ \citenamefont {Wiedemann}}]{Schwarz2015}%
  \BibitemOpen
  \bibfield  {author} {\bibinfo {author} {\bibfnamefont {M.}~\bibnamefont
  {Schwarz}}, \bibinfo {author} {\bibfnamefont {E.-A.}\ \bibnamefont {Knabbe}},
  \bibinfo {author} {\bibfnamefont {A.}~\bibnamefont {Lindner}}, \bibinfo
  {author} {\bibfnamefont {J.}~\bibnamefont {Redondo}}, \bibinfo {author}
  {\bibfnamefont {A.}~\bibnamefont {Ringwald}}, \bibinfo {author}
  {\bibfnamefont {M.}~\bibnamefont {Schneide}}, \bibinfo {author}
  {\bibfnamefont {J.}~\bibnamefont {Susol}}, \ and\ \bibinfo {author}
  {\bibfnamefont {G.}~\bibnamefont {Wiedemann}},\ }\href {\doibase
  10.1088/1475-7516/2015/08/011} {\bibfield  {journal} {\bibinfo  {journal} {J.
  Cosmol. Astropart. Phys.}\ ,\ \bibinfo {pages} {011}} (\bibinfo {year}
  {2015})}\BibitemShut {NoStop}%
\bibitem [{\citenamefont {Redondo}(2008)}]{Redondo2008}%
  \BibitemOpen
  \bibfield  {author} {\bibinfo {author} {\bibfnamefont {J.}~\bibnamefont
  {Redondo}},\ }\href {\doibase 10.1088/1475-7516/2008/07/008} {\bibfield
  {journal} {\bibinfo  {journal} {J. Cosmol. Astropart. Phys.}\ ,\ \bibinfo
  {pages} {008}} (\bibinfo {year} {2008})}\BibitemShut {NoStop}%
\bibitem [{\citenamefont {An}\ \emph {et~al.}(2013{\natexlab{a}})\citenamefont
  {An}, \citenamefont {Pospelov},\ and\ \citenamefont {Pradler}}]{An2013New}%
  \BibitemOpen
  \bibfield  {author} {\bibinfo {author} {\bibfnamefont {H.}~\bibnamefont
  {An}}, \bibinfo {author} {\bibfnamefont {M.}~\bibnamefont {Pospelov}}, \ and\
  \bibinfo {author} {\bibfnamefont {J.}~\bibnamefont {Pradler}},\ }\href
  {\doibase 10.1016/j.physletb.2013.07.008} {\bibfield  {journal} {\bibinfo
  {journal} {Phys. Lett. B}\ }\textbf {\bibinfo {volume} {725}},\ \bibinfo
  {pages} {190 } (\bibinfo {year} {2013}{\natexlab{a}})}\BibitemShut {NoStop}%
\bibitem [{\citenamefont {Mirizzi}\ \emph {et~al.}(2009)\citenamefont
  {Mirizzi}, \citenamefont {Redondo},\ and\ \citenamefont
  {Sigl}}]{Mirizzi2009}%
  \BibitemOpen
  \bibfield  {author} {\bibinfo {author} {\bibfnamefont {A.}~\bibnamefont
  {Mirizzi}}, \bibinfo {author} {\bibfnamefont {J.}~\bibnamefont {Redondo}}, \
  and\ \bibinfo {author} {\bibfnamefont {G.}~\bibnamefont {Sigl}},\ }\href
  {\doibase 10.1088/1475-7516/2009/03/026} {\bibfield  {journal} {\bibinfo
  {journal} {J. Cosmol. Astropart. Phys.}\ ,\ \bibinfo {pages} {026}} (\bibinfo
  {year} {2009})}\BibitemShut {NoStop}%
\bibitem [{\citenamefont {Aguilar-Arevalo}\ \emph {et~al.}(2019)\citenamefont
  {Aguilar-Arevalo} \emph {et~al.}}]{Aguilar2019}%
  \BibitemOpen
  \bibfield  {author} {\bibinfo {author} {\bibfnamefont {A.}~\bibnamefont
  {Aguilar-Arevalo}} \emph {et~al.} (\bibinfo {collaboration} {{DAMIC
  Collaboration}}),\ }\href {\doibase 10.1103/PhysRevLett.123.181802}
  {\bibfield  {journal} {\bibinfo  {journal} {Phys. Rev. Lett.}\ }\textbf
  {\bibinfo {volume} {123}},\ \bibinfo {pages} {181802} (\bibinfo {year}
  {2019})}\BibitemShut {NoStop}%
\bibitem [{\citenamefont {Angloher}\ \emph {et~al.}(2017)\citenamefont
  {Angloher} \emph {et~al.}}]{Angloher2017}%
  \BibitemOpen
  \bibfield  {author} {\bibinfo {author} {\bibfnamefont {G.}~\bibnamefont
  {Angloher}} \emph {et~al.},\ }\href {\doibase 10.1140/epjc/s10052-017-4878-6}
  {\bibfield  {journal} {\bibinfo  {journal} {Eur. Phys. J. C}\ }\textbf
  {\bibinfo {volume} {77}},\ \bibinfo {pages} {299} (\bibinfo {year}
  {2017})}\BibitemShut {NoStop}%
\bibitem [{\citenamefont {Gninenko}\ and\ \citenamefont
  {Redondo}(2008)}]{Gninenko2008}%
  \BibitemOpen
  \bibfield  {author} {\bibinfo {author} {\bibfnamefont {S.~N.}\ \bibnamefont
  {Gninenko}}\ and\ \bibinfo {author} {\bibfnamefont {J.}~\bibnamefont
  {Redondo}},\ }\href {\doibase 10.1016/j.physletb.2008.05.020} {\bibfield
  {journal} {\bibinfo  {journal} {Phys. Lett. B}\ }\textbf {\bibinfo {volume}
  {664}},\ \bibinfo {pages} {180} (\bibinfo {year} {2008})}\BibitemShut
  {NoStop}%
\bibitem [{\citenamefont {An}\ \emph {et~al.}(2015)\citenamefont {An},
  \citenamefont {Pospelov}, \citenamefont {Pradler},\ and\ \citenamefont
  {Ritz}}]{An2015}%
  \BibitemOpen
  \bibfield  {author} {\bibinfo {author} {\bibfnamefont {H.}~\bibnamefont
  {An}}, \bibinfo {author} {\bibfnamefont {M.}~\bibnamefont {Pospelov}},
  \bibinfo {author} {\bibfnamefont {J.}~\bibnamefont {Pradler}}, \ and\
  \bibinfo {author} {\bibfnamefont {A.}~\bibnamefont {Ritz}},\ }\href {\doibase
  10.1016/j.physletb.2015.06.018} {\bibfield  {journal} {\bibinfo  {journal}
  {Phys. Lett. B}\ }\textbf {\bibinfo {volume} {747}},\ \bibinfo {pages} {331 }
  (\bibinfo {year} {2015})}\BibitemShut {NoStop}%
\bibitem [{\citenamefont {An}\ \emph {et~al.}(2013{\natexlab{b}})\citenamefont
  {An}, \citenamefont {Pospelov},\ and\ \citenamefont {Pradler}}]{An2013Dark}%
  \BibitemOpen
  \bibfield  {author} {\bibinfo {author} {\bibfnamefont {H.}~\bibnamefont
  {An}}, \bibinfo {author} {\bibfnamefont {M.}~\bibnamefont {Pospelov}}, \ and\
  \bibinfo {author} {\bibfnamefont {J.}~\bibnamefont {Pradler}},\ }\href
  {\doibase 10.1103/PhysRevLett.111.041302} {\bibfield  {journal} {\bibinfo
  {journal} {Phys. Rev. Lett.}\ }\textbf {\bibinfo {volume} {111}},\ \bibinfo
  {pages} {041302} (\bibinfo {year} {2013}{\natexlab{b}})}\BibitemShut
  {NoStop}%
\bibitem [{\citenamefont {Bloch}\ \emph {et~al.}(2017)\citenamefont {Bloch},
  \citenamefont {Essig}, \citenamefont {Tobioka}, \citenamefont {Volansky},\
  and\ \citenamefont {Yu}}]{Bloch2017}%
  \BibitemOpen
  \bibfield  {author} {\bibinfo {author} {\bibfnamefont {I.~M.}\ \bibnamefont
  {Bloch}}, \bibinfo {author} {\bibfnamefont {R.}~\bibnamefont {Essig}},
  \bibinfo {author} {\bibfnamefont {K.}~\bibnamefont {Tobioka}}, \bibinfo
  {author} {\bibfnamefont {T.}~\bibnamefont {Volansky}}, \ and\ \bibinfo
  {author} {\bibfnamefont {T.-T.}\ \bibnamefont {Yu}},\ }\href {\doibase
  10.1007/JHEP06(2017)087} {\bibfield  {journal} {\bibinfo  {journal} {J. High
  Energy Phys.}\ ,\ \bibinfo {pages} {87}} (\bibinfo {year}
  {2017})}\BibitemShut {NoStop}%
\bibitem [{\citenamefont {Aprile}\ \emph {et~al.}(2019)\citenamefont {Aprile}
  \emph {et~al.}}]{XENON1t}%
  \BibitemOpen
  \bibfield  {author} {\bibinfo {author} {\bibfnamefont {E.}~\bibnamefont
  {Aprile}} \emph {et~al.} (\bibinfo {collaboration} {XENON Collaboration}),\
  }\href {\doibase 10.1103/PhysRevLett.123.251801} {\bibfield  {journal}
  {\bibinfo  {journal} {Phys. Rev. Lett.}\ }\textbf {\bibinfo {volume} {123}},\
  \bibinfo {pages} {251801} (\bibinfo {year} {2019})}\BibitemShut {NoStop}%
\bibitem [{\citenamefont {Abramoff}\ \emph {et~al.}(2019)\citenamefont
  {Abramoff} \emph {et~al.}}]{Abramoff2019}%
  \BibitemOpen
  \bibfield  {author} {\bibinfo {author} {\bibfnamefont {O.}~\bibnamefont
  {Abramoff}} \emph {et~al.} (\bibinfo {collaboration} {SENSEI
  Collaboration}),\ }\href {\doibase 10.1103/PhysRevLett.122.161801} {\bibfield
   {journal} {\bibinfo  {journal} {Phys. Rev. Lett.}\ }\textbf {\bibinfo
  {volume} {122}},\ \bibinfo {pages} {161801} (\bibinfo {year}
  {2019})}\BibitemShut {NoStop}%
\bibitem [{\citenamefont {Kang}\ \emph {et~al.}(2013)\citenamefont {Kang} \emph
  {et~al.}}]{Kang2013}%
  \BibitemOpen
  \bibfield  {author} {\bibinfo {author} {\bibfnamefont {K.~J.}\ \bibnamefont
  {Kang}} \emph {et~al.} (\bibinfo {collaboration} {CDEX Collaboration}),\
  }\href {\doibase 10.1007/s11467-013-0349-1} {\bibfield  {journal} {\bibinfo
  {journal} {Front. Phys.}\ }\textbf {\bibinfo {volume} {8}},\ \bibinfo {pages}
  {412} (\bibinfo {year} {2013})}\BibitemShut {NoStop}%
\bibitem [{\citenamefont {Cheng}\ \emph {et~al.}(2017)\citenamefont {Cheng}
  \emph {et~al.}}]{Cheng2017}%
  \BibitemOpen
  \bibfield  {author} {\bibinfo {author} {\bibfnamefont {J.~P.}\ \bibnamefont
  {Cheng}} \emph {et~al.},\ }\href {\doibase
  10.1146/annurev-nucl-102115-044842} {\bibfield  {journal} {\bibinfo
  {journal} {Annu. Rev. Nucl. Part. Sci.}\ }\textbf {\bibinfo {volume} {67}},\
  \bibinfo {pages} {231} (\bibinfo {year} {2017})}\BibitemShut {NoStop}%
\bibitem [{\citenamefont {Wu}\ \emph {et~al.}(2013)\citenamefont {Wu} \emph
  {et~al.}}]{Wu2013}%
  \BibitemOpen
  \bibfield  {author} {\bibinfo {author} {\bibfnamefont {Y.~C.}\ \bibnamefont
  {Wu}} \emph {et~al.} (\bibinfo {collaboration} {CDEX Collaboration}),\ }\href
  {\doibase 10.1088/1674-1137/37/8/086001} {\bibfield  {journal} {\bibinfo
  {journal} {Chin. Phys. C}\ }\textbf {\bibinfo {volume} {37}},\ \bibinfo
  {pages} {086001} (\bibinfo {year} {2013})}\BibitemShut {NoStop}%
\bibitem [{\citenamefont {Yang}\ \emph {et~al.}(2019)\citenamefont {Yang} \emph
  {et~al.}}]{Yang2019}%
  \BibitemOpen
  \bibfield  {author} {\bibinfo {author} {\bibfnamefont {L.~T.}\ \bibnamefont
  {Yang}} \emph {et~al.} (\bibinfo {collaboration} {CDEX Collaboration}),\
  }\href {\doibase 10.1103/PhysRevLett.123.221301} {\bibfield  {journal}
  {\bibinfo  {journal} {Phys. Rev. Lett.}\ }\textbf {\bibinfo {volume} {123}},\
  \bibinfo {pages} {221301} (\bibinfo {year} {2019})}\BibitemShut {NoStop}%
\bibitem [{\citenamefont {Liu}\ \emph {et~al.}(2019)\citenamefont {Liu} \emph
  {et~al.}}]{Liu2019}%
  \BibitemOpen
  \bibfield  {author} {\bibinfo {author} {\bibfnamefont {Z.~Z.}\ \bibnamefont
  {Liu}} \emph {et~al.} (\bibinfo {collaboration} {CDEX Collaboration}),\
  }\href {\doibase 10.1103/PhysRevLett.123.161301} {\bibfield  {journal}
  {\bibinfo  {journal} {Phys. Rev. Lett}\ }\textbf {\bibinfo {volume} {123}},\
  \bibinfo {pages} {161301} (\bibinfo {year} {2019})}\BibitemShut {NoStop}%
\bibitem [{\citenamefont {Jiang}\ \emph {et~al.}(2018)\citenamefont {Jiang}
  \emph {et~al.}}]{Jiang2018}%
  \BibitemOpen
  \bibfield  {author} {\bibinfo {author} {\bibfnamefont {H.}~\bibnamefont
  {Jiang}} \emph {et~al.} (\bibinfo {collaboration} {CDEX Collaboration}),\
  }\href {\doibase 10.1103/PhysRevLett.120.241301} {\bibfield  {journal}
  {\bibinfo  {journal} {Phys. Rev. Lett.}\ }\textbf {\bibinfo {volume} {120}},\
  \bibinfo {pages} {241301} (\bibinfo {year} {2018})}\BibitemShut {NoStop}%
\bibitem [{\citenamefont {Jiang}\ \emph {et~al.}(2019)\citenamefont {Jiang}
  \emph {et~al.}}]{Jiang2019}%
  \BibitemOpen
  \bibfield  {author} {\bibinfo {author} {\bibfnamefont {H.}~\bibnamefont
  {Jiang}} \emph {et~al.} (\bibinfo {collaboration} {CDEX Collaboration}),\
  }\href {\doibase 10.1007/s11433-018-8001-3} {\bibfield  {journal} {\bibinfo
  {journal} {Sci. China-Phys. Mech. Astron.}\ }\textbf {\bibinfo {volume}
  {62}},\ \bibinfo {pages} {31012} (\bibinfo {year} {2019})}\BibitemShut
  {NoStop}%
\bibitem [{\citenamefont {Yang}\ \emph
  {et~al.}(2018{\natexlab{a}})\citenamefont {Yang} \emph
  {et~al.}}]{Yang2018Bulk}%
  \BibitemOpen
  \bibfield  {author} {\bibinfo {author} {\bibfnamefont {L.~T.}\ \bibnamefont
  {Yang}} \emph {et~al.},\ }\href {\doibase 10.1016/j.nima.2017.12.078}
  {\bibfield  {journal} {\bibinfo  {journal} {Nucl. Instrum. Methods Phys.
  Res., Sect A}\ }\textbf {\bibinfo {volume} {886}},\ \bibinfo {pages} {13}
  (\bibinfo {year} {2018}{\natexlab{a}})}\BibitemShut {NoStop}%
\bibitem [{\citenamefont {Yang}\ \emph
  {et~al.}(2018{\natexlab{b}})\citenamefont {Yang} \emph {et~al.}}]{Yang2018}%
  \BibitemOpen
  \bibfield  {author} {\bibinfo {author} {\bibfnamefont {L.~T.}\ \bibnamefont
  {Yang}} \emph {et~al.} (\bibinfo {collaboration} {CDEX Collaboration}),\
  }\href {\doibase 10.1088/1674-1137/42/2/023002} {\bibfield  {journal}
  {\bibinfo  {journal} {Chin. Phys. C}\ }\textbf {\bibinfo {volume} {42}},\
  \bibinfo {pages} {023002} (\bibinfo {year} {2018}{\natexlab{b}})}\BibitemShut
  {NoStop}%
\bibitem [{\citenamefont {Yue}\ \emph {et~al.}(2014)\citenamefont {Yue} \emph
  {et~al.}}]{Yue2014}%
  \BibitemOpen
  \bibfield  {author} {\bibinfo {author} {\bibfnamefont {Q.}~\bibnamefont
  {Yue}} \emph {et~al.} (\bibinfo {collaboration} {CDEX Collaboration}),\
  }\href {\doibase 10.1103/PhysRevD.90.091701} {\bibfield  {journal} {\bibinfo
  {journal} {Phys. Rev. D}\ }\textbf {\bibinfo {volume} {90}},\ \bibinfo
  {pages} {091701} (\bibinfo {year} {2014})}\BibitemShut {NoStop}%
\bibitem [{\citenamefont {Bahcall}(1963)}]{Bahcall1963}%
  \BibitemOpen
  \bibfield  {author} {\bibinfo {author} {\bibfnamefont {J.~N.}\ \bibnamefont
  {Bahcall}},\ }\href {\doibase 10.1103/PhysRev.132.362} {\bibfield  {journal}
  {\bibinfo  {journal} {Phys. Rev.}\ }\textbf {\bibinfo {volume} {132}},\
  \bibinfo {pages} {362} (\bibinfo {year} {1963})}\BibitemShut {NoStop}%
\bibitem [{\citenamefont {Feldman}\ and\ \citenamefont
  {Cousins}(1998)}]{Feldman1998}%
  \BibitemOpen
  \bibfield  {author} {\bibinfo {author} {\bibfnamefont {G.~J.}\ \bibnamefont
  {Feldman}}\ and\ \bibinfo {author} {\bibfnamefont {R.~D.}\ \bibnamefont
  {Cousins}},\ }\href {\doibase 10.1103/PhysRevD.57.3873} {\bibfield  {journal}
  {\bibinfo  {journal} {Phys. Rev. D}\ }\textbf {\bibinfo {volume} {57}},\
  \bibinfo {pages} {3873} (\bibinfo {year} {1998})}\BibitemShut {NoStop}%
\bibitem [{\citenamefont {Horvat}\ \emph {et~al.}(2013)\citenamefont {Horvat},
  \citenamefont {Kekez}, \citenamefont {M.Krčmar}, \citenamefont {Krečak},\
  and\ \citenamefont {Ljubičić}}]{Horvat2013}%
  \BibitemOpen
  \bibfield  {author} {\bibinfo {author} {\bibfnamefont {R.}~\bibnamefont
  {Horvat}}, \bibinfo {author} {\bibfnamefont {D.}~\bibnamefont {Kekez}},
  \bibinfo {author} {\bibnamefont {M.Krčmar}}, \bibinfo {author}
  {\bibfnamefont {Z.}~\bibnamefont {Krečak}}, \ and\ \bibinfo {author}
  {\bibfnamefont {A.}~\bibnamefont {Ljubičić}},\ }\href {\doibase
  10.1016/j.physletb.2013.03.014} {\bibfield  {journal} {\bibinfo  {journal}
  {Phys. Lett. B}\ }\textbf {\bibinfo {volume} {721}},\ \bibinfo {pages} {220 }
  (\bibinfo {year} {2013})}\BibitemShut {NoStop}%
\bibitem [{\citenamefont {Henke}\ \emph {et~al.}(1993)\citenamefont {Henke},
  \citenamefont {Gullikson},\ and\ \citenamefont {Davis}}]{Henke1993}%
  \BibitemOpen
  \bibfield  {author} {\bibinfo {author} {\bibfnamefont {B.~L.}\ \bibnamefont
  {Henke}}, \bibinfo {author} {\bibfnamefont {E.~M.}\ \bibnamefont
  {Gullikson}}, \ and\ \bibinfo {author} {\bibfnamefont {J.~C.}\ \bibnamefont
  {Davis}},\ }\href {\doibase 10.1016/0092-640X(82)90002-X} {\bibfield
  {journal} {\bibinfo  {journal} {At. Data Nucl. Data Tables}\ }\textbf
  {\bibinfo {volume} {54}},\ \bibinfo {pages} {181 } (\bibinfo {year}
  {1993})}\BibitemShut {NoStop}%
\bibitem [{\citenamefont {Gondolo}\ and\ \citenamefont
  {Raffelt}(2009)}]{Gondolo2009}%
  \BibitemOpen
  \bibfield  {author} {\bibinfo {author} {\bibfnamefont {P.}~\bibnamefont
  {Gondolo}}\ and\ \bibinfo {author} {\bibfnamefont {G.~G.}\ \bibnamefont
  {Raffelt}},\ }\href {\doibase 10.1103/PhysRevD.79.107301} {\bibfield
  {journal} {\bibinfo  {journal} {Phys. Rev. D}\ }\textbf {\bibinfo {volume}
  {79}},\ \bibinfo {pages} {107301} (\bibinfo {year} {2009})}\BibitemShut
  {NoStop}%
\bibitem [{\citenamefont {Redondo}\ and\ \citenamefont
  {Raffelt}(2013)}]{Redondo2013}%
  \BibitemOpen
  \bibfield  {author} {\bibinfo {author} {\bibfnamefont {J.}~\bibnamefont
  {Redondo}}\ and\ \bibinfo {author} {\bibfnamefont {G.}~\bibnamefont
  {Raffelt}},\ }\href {\doibase 10.1088/1475-7516/2013/08/034} {\bibfield
  {journal} {\bibinfo  {journal} {J. Cosmol. Astropart. Phys.}\ ,\ \bibinfo
  {pages} {034}} (\bibinfo {year} {2013})}\BibitemShut {NoStop}%
\bibitem [{\citenamefont {Aalseth}\ \emph {et~al.}(2011)\citenamefont {Aalseth}
  \emph {et~al.}}]{Aalseth2011}%
  \BibitemOpen
  \bibfield  {author} {\bibinfo {author} {\bibfnamefont {C.~E.}\ \bibnamefont
  {Aalseth}} \emph {et~al.} (\bibinfo {collaboration} {CoGeNT Collaboration}),\
  }\href {\doibase 10.1103/PhysRevLett.106.131301} {\bibfield  {journal}
  {\bibinfo  {journal} {Phys. Rev. Lett.}\ }\textbf {\bibinfo {volume} {106}},\
  \bibinfo {pages} {131301} (\bibinfo {year} {2011})}\BibitemShut {NoStop}%
\bibitem [{\citenamefont {Navarro}\ \emph {et~al.}(1997)\citenamefont
  {Navarro}, \citenamefont {Frenk},\ and\ \citenamefont {White}}]{Navarro1996}%
  \BibitemOpen
  \bibfield  {author} {\bibinfo {author} {\bibfnamefont {J.~F.}\ \bibnamefont
  {Navarro}}, \bibinfo {author} {\bibfnamefont {C.~S.}\ \bibnamefont {Frenk}},
  \ and\ \bibinfo {author} {\bibfnamefont {S.~D.~M.}\ \bibnamefont {White}},\
  }\href {\doibase 10.1086/304888} {\bibfield  {journal} {\bibinfo  {journal}
  {Astrophys. J}\ }\textbf {\bibinfo {volume} {490}},\ \bibinfo {pages} {493}
  (\bibinfo {year} {1997})}\BibitemShut {NoStop}%
\end{thebibliography}%
\end{document}